\documentclass[aps,prd,english,showpacs,11pt]{revtex4-1}
\usepackage[latin1]{inputenc}
\usepackage{ucs}
\usepackage{amsmath}
\usepackage{amsfonts}
\usepackage{amssymb}
\usepackage{graphicx}
\usepackage{psfrag}
\usepackage{picinpar}
\usepackage{hyperref}
\usepackage{color}
\hypersetup{ 
     colorlinks   = true,
     citecolor    = blue,
     linkcolor    = darkgreen,
     urlcolor     = black
}

\allowdisplaybreaks[1]
\newcommand{\be}{\begin{equation}}
\newcommand{\ee}{\end{equation}}
\newcommand{\ben}{\begin{eqnarray}}
\newcommand{\een}{\end{eqnarray}}
\newcommand{\bes}{\begin{subequations}}
\newcommand{\ees}{\end{subequations}}

\newcommand{\Frac}[2]{\frac{{\displaystyle #1}}{{\displaystyle #2}}}
\definecolor{darkgreen}{rgb}{0,0.5,0}

\begin{document}
\title{Constraining condensate dark matter in galaxy clusters}

\author{J. C. C. de Souza} 
\email{jose.souza@ufabc.edu.br}
\author{M. Ujevic}
\email{mujevic@ufabc.edu.br}

\affiliation{Centro de Ci\^{e}ncias Naturais e Humanas, Universidade
  Federal do ABC, Rua Santa Ad\'elia 166, 09210-170, Santo Andr\'{e}, SP,
  Brazil}

\begin{abstract}
We constrain scattering length parameters in a Bose-Einstein
condensate dark matter model by using galaxy clusters radii, with the
implementation of a method previously applied to galaxies. At the
present work, we use a sample of 114 clusters radii in order to obtain
the scattering lengths associated with a dark matter particle mass in
the range $10^{-6}-10^{-4}\, {\rm eV}$. We obtain scattering lengths
that are five orders of magnitude larger than the ones found in the
galactic case, even when taking into account the cosmological
expansion in the cluster scale by means of the introduction of a small
cosmological constant. We also construct and compare curves for the
orbital velocity of a test particle in the vicinity of a dark matter
cluster in both the expanding and the non-expanding cases.

\end{abstract}


\maketitle

\section{Introduction}

 It has long been observed that almost $27\%$ of the energy
 density in the Universe is in the form of a rather mysterious entity
 dubbed dark matter \cite{wmap, planck}. So far, investigations on the
 nature of this sort of matter have presented no definitive
 conclusions.
 
 When it comes to the dark matter present in structures such as
 galaxies and clusters, many proposals have been put forward. We can
 mention {\it Weakly Interacting Massive Particles} (WIMP's) as one of
 the most popular suggestions. These particles are very massive
 ($\mathcal{O}({\rm GeV})$) and present very small coupling constants
 to the baryonic matter \cite{pdg12}. Experimental searches for WIMP's
 are being presently performed.

 Recently, it has been proposed that the dark matter composing
 structures in the Universe is in the form of a Bose-Einstein
 condensate (BEC) \cite{sik09,boh07,chavanis,chavanis2} which suffered
 a gravitational collapse \cite{khlopov85, suarez13}. This condensate
 allows for the construction of suitable rotation curves for galaxies.
 The particles suggested to compose this condensate are sub-eV in mass
 and have little interaction with baryonic matter. They have been
 grouped in the {\it Weakly Interacting Slim Particles} (WISP's)
 \cite{arias} category, which includes the QCD axion \cite{peccei,
   peccei2} as the most prominent member.

 Considering axion-like particles (particles in the mass range of the
 axion but with the possibility of having spin 0 or 1), the parameters
 of the condensate have been constrained by using galaxies' radii data
 \cite{pir12}. In the present work, we apply the same procedure to
 galaxy clusters. We use a data set of 114 clusters radii and perform
 a statistical analysis to obtain the most representative scattering
 length value for a specific particle mass.
 Moreover, since cluster sizes may be in a scale which can be affected
 by the cosmological expansion, we repeat this procedure including a
 small valued cosmological constant in a Newtonian approximation
 \cite{nandra0,nandra}, in the attempt to obtain noticeably distinct results
 from the non-expanding case (for a relativistic version of BEC dark
 matter, see, e.g, \cite{bettoni14}).

 This paper is organized as follows: section \ref{BEC} presents a
 brief review of the theoretical background on Bose-Einstein
 condensate dark matter. Section \ref{rho_clus} shows the density
 profiles in both static and expanding cases along with the
 corresponding cluster radii. In section \ref{stats}, we perform the
 statistical analysis that allows us to obtain the scattering lengths
 for the dark matter particle, in the mass range $10^{-6}-10^{-4}\,
 {\rm eV}$. We show the orbital velocity of a test particle under the
 influence of the cluster mass in section \ref{vel_clus}. Our
 conclusions appear in section \ref{conc}.

\section{Bose-Einstein condensate dark matter}\label{BEC}

 We recall here the theoretical description of the galactic
 Bose-Einstein condensate composed of axionlike particles.
 
 At zero temperature, the dynamics of the field destruction operator
 $\hat{\psi}({\bf r},t)$ (representing each dark matter particle) in
the Heisenberg picture, $-i\hslash\partial_t\hat{\psi}({\bf
  r},t)=[\hat{H},\hat{\psi}({\bf r},t)]$, yields the time-independent
Gross-Pitaevskii equation (GPE) for the BEC wavefunction $\psi({\bf
  r})$ \cite{pethick} 
\be\label{GPEI} \mu \psi({\bf
  r})=-\Frac{\hslash^2}{2m}\nabla^{2}\psi+V({\bf r})\psi({\bf
  r})+\Frac{4\pi \hslash^2 a}{m}|\psi({\bf r})|^{2}\psi({\bf r})\; ,
\ee
\noindent where $m$ is the mass of the particle, $a$ is the $s$-wave
scattering length which characterizes two-body collisions between
particles, $V({\bf r})$ is the trapping potential and $\mu$ is the
chemical potential.

When the potential $V({\bf r})$ obeys the Poisson's equation (which is
the case of a self-gravitating condensate),
\begin{eqnarray}\label{poisson}
\nabla^2V=4\pi G m\rho_{\!_{DM}}\;  ,
\end{eqnarray}
\noindent where $\rho_{\!_{DM}}$ is the particle number density of the dark
matter concentration, and we consider a large number of particles
(this is the Thomas-Fermi (TF) approximation \cite{souza14}), it has been
demonstrated \cite{boh07, pir12} that (\ref{GPEI}) has the solution
\be\label{psibh} 
|\psi_{BH}(r)|^{2}=\rho(r)=\begin{cases} \rho_0\Frac{\sin
    kr}{kr}\quad\mbox{for}\quad r\le R \\ 0\quad \mbox{for} \quad
r>R \end{cases}\; , 
\ee
\noindent with $k=\sqrt{Gm^3/\hslash^2a}$, $R=\pi/k$ and $\rho_{0}$ is
the central particle number density of the condensate. This is the
Boehmer-Harko (BF) solution. It results in a halo radius given by
\be\label{radius0}
R=\pi \sqrt{\Frac{\hslash^{2} a}{G m^{3}}}\;.
\ee
\noindent Using this relation and considering the dark matter particle mass
range $10^{-6}-10^{-4}\; {\rm eV}$, the lower bound of the scattering length has been
constrained to $10^{-29}\ {\rm m}$ in galaxies \cite{pir12}.

 The same results apply to the case of a particle with spin-1, with
 the important difference that now the condensate may assume two
 distinct states, polar (when the particles spins are antiparallel)
 and ferromagnetic (parallel spins) \cite{kaw11}. 

 Hence, for the polar state, the radius is given by
 \begin{equation}\label{rpolar}
 R_p = \pi \sqrt{\frac{\hbar^2 (a^p_0+2a^p_2)}{3Gm^3}}\, ,
\end{equation}
\noindent where $a^p_0$ and $a^p_2$ are the scattering lenghts related
to this phase.

 For the ferromagnetic phase, one obtains 
 \begin{equation}\label{rferr}
 R_f = \pi \sqrt{\frac{\hbar^2 a^f_2}{Gm^3}}\;, 
 \end{equation}
 \noindent where $a^f_2$ is a new scattering length for this particular
 state.

The central mass density $\varrho_{\!_{0}}=m\rho_{\!_{0}}$ will be assumed
throughout this paper to be the one of a typical cluster with mass $M\sim
10^{14}\,M_{\odot}$ and radius $R\sim 1\, {\rm Mpc}$, yielding
$\varrho_{\!_{0}}\approx 10^{-24}{\rm \, kg\, m^{-3}}$.

\section{Density profile and radii of galaxy clusters with a
  cosmological constant}\label{rho_clus}
 
  Following the assumptions made in \cite{nandra0,nandra}, we can
  consider the cluster to be embedded in an expanding spacetime
  background, with the expansion rate given by the Hubble parameter
  $H=\sqrt{\Lambda/3}$ and $\Lambda$ being a small cosmological
  constant.

 For the purpose of describing the cluster dark matter as a
 condensate, the effect of the expansion is equivalent to the
 introduction of an additional repulsive radial potential
 \be\label{potlambda} V_{\!_{\Lambda}}(r)=-\Frac{m}{6}\Lambda r^{2}
 \ee
 \noindent in the GPE (\ref{GPEI}). The total potential which confines the cluster is now
 \be
V'(r)=V(r)+V_{\!_{\Lambda}}(r)\; ,
 \ee
\noindent where $V(r)$ is the gravitational potential. When  $V'(r)$ obeys the Poisson equation
\be\label{poissonlambda}
\nabla^2V'=4\pi G m\rho_{\!_{\Lambda}}\;,
\ee
\noindent where $\rho_{\!_{\Lambda}}$ is the particle density in the
presence of the cosmological constant, differentiation of equation
(\ref{GPEI}) results in
\be\label{prelane}
2\pi G m\rho-m\Lambda+\Frac{2\pi \hslash^2 a}{m}\nabla^2\rho=0\;.
\ee
\noindent With the use of the identification
$\rho_{\!_{\Lambda}}=\Frac{m^2}{2\pi \hslash^2 a}(2\pi G
\rho-\Lambda)$, equation (\ref{prelane}) can be recast in the form
of the usual Lane-Emden equation
\be\label{lane-emden}
\frac{1}{\xi^2}\frac{\partial}{\partial\xi}\left(\xi^2
\frac{\partial\theta}{\partial\xi}\right)+\theta^n=0\;,
\ee
\noindent where we have used  $\rho_{\!_{\Lambda}} =
\rho_{\!_{0}}\theta^n$, and $\theta$ being a
function of the dimensionless coordinate $\xi$ defined by $r =
[(n+1)K\rho_c^{1/n-1}/4\pi G]^{1/2}\xi$. We recall that for a static
condensate we have the polytropic equation of state, relating the
density and the density and the pressure of the fluid, $p =
K\varrho_{\!_{\Lambda}}^{1+\frac{1}{n}}$, with $K$ a constant and $n$ the polytropic
index.

 In the present case, $n=1$ and $K=2\pi\hbar^2a/m^3$, making it possible
 to obtain the analytical solution for the Lane-Emden equation as
\begin{equation}
\theta(\xi) = \frac{\sin(\xi)}{\xi}\;.
\end{equation}

\begin{figure}[!htp]
\begin{center}

\includegraphics[scale=0.63]{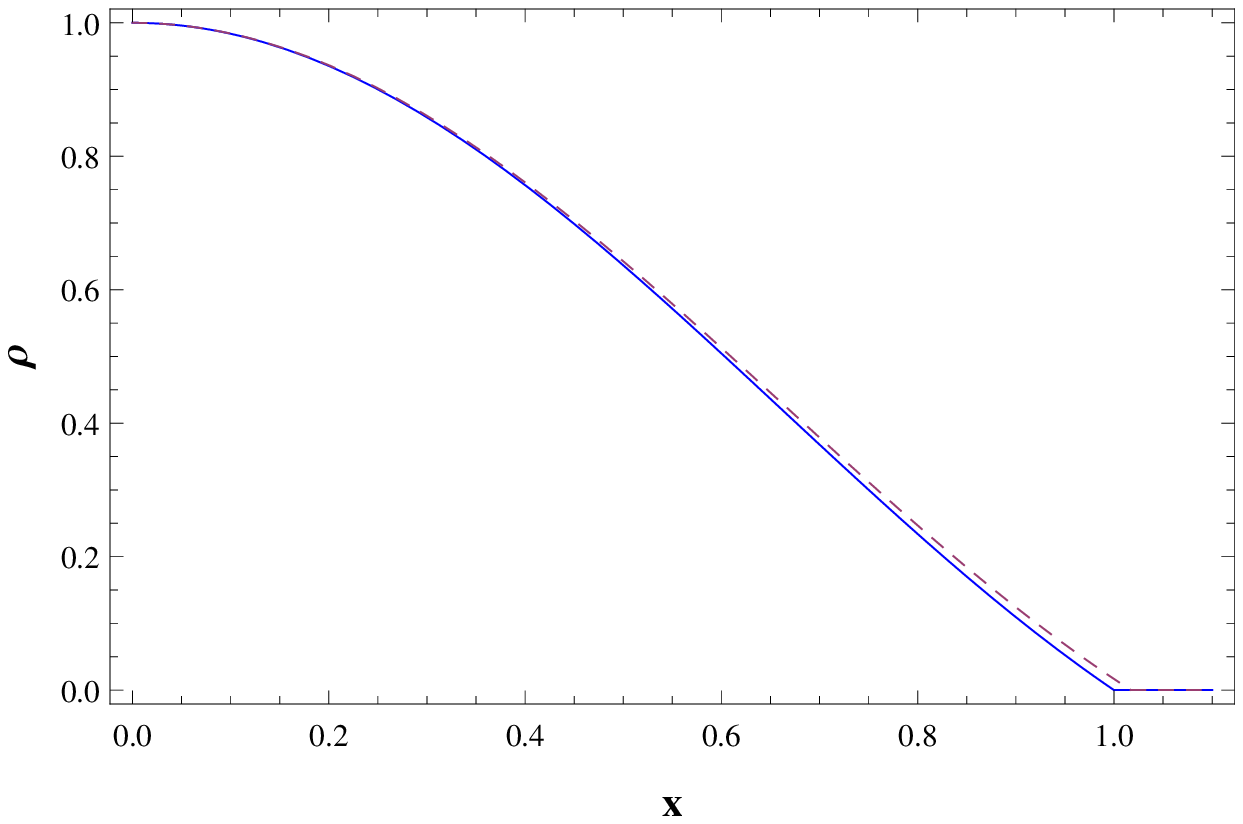}
\hspace{0.1cm}
\includegraphics[scale=0.63]{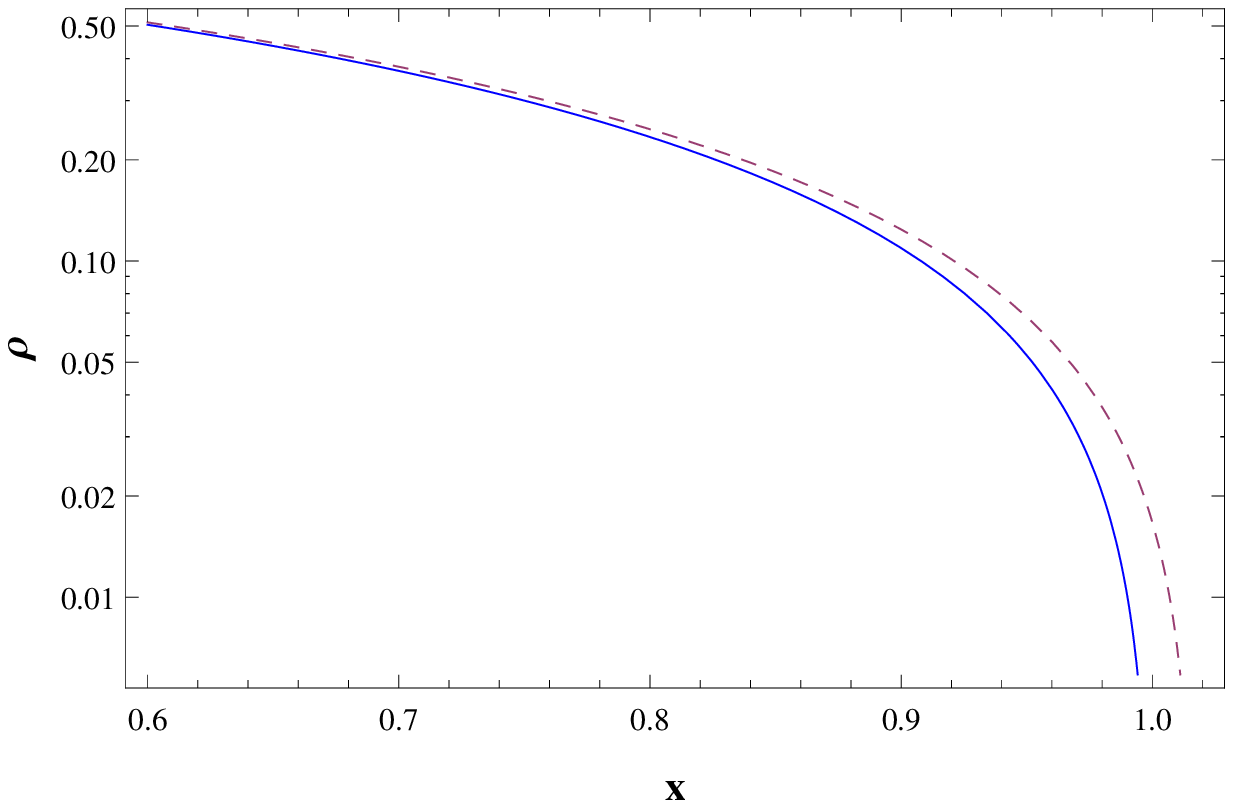}
\caption{Particle number density profile $\rho_{BH}$ (in units of $(\rho_{0}/\pi)$)
  corresponding to the Boehmer-Harko solution in the Thomas-Fermi
  approximation with and without a cosmological constant (dashed and
  solid lines, respectively). The right panel shows a zoom in
  logarithmic scale in order to stress the difference between the
  functions. In both cases $x=r/R$. A central cluster mass density
  $\varrho_{\!_{0}}=10^{-24}\, {\rm kg\, m^{-3}}$ and a cosmological constant
  $\Lambda=1.4\times 10^{-35}\, {\rm s^{-2}}$ \cite{planck} have been used
  in both plots.}
\label{rho_lambda}

\end{center}
\end{figure}

 With the appropriate boundary conditions for the condensate, the
 particle number density profile obtained from (\ref{GPEI}) and
 (\ref{lane-emden}) for a cluster under the action of a cosmological
 constant is thus
 \be\label{rholambda}
 \rho_{\!_{\Lambda}}(r)=\begin{cases} \left(\rho_0-\Frac{\Lambda}{4\pi
   m G}\right)\Frac{\sin (kr)}{kr}+\Frac{\Lambda}{4\pi m
   G}\quad\mbox{for}\quad r\le \bar{R} \\ 0\quad \mbox{for} \quad
 r>\bar{R} \end{cases}\;,
 \ee
 \noindent where $\bar{R}$ is the cluster radius in the expanding
 environment. We can see that the density has been rescaled and shifted
 by a small quantity that depends on the magnitude of $\Lambda$. For
 the sake of comparison, the density profiles in the expanding and
 non-expanding cases are depicted in figure \ref{rho_lambda}.

 For a spherically symmetric mass distribution, the total 
 mass inside a radius $r$, resulting from (\ref{rholambda}), is calculated as
 \be\label{masslamb}
 M_{\!_{\Lambda}}(r)=4 \pi m
 \int_{0}^{r}\rho_{\!_{\Lambda}}(r')(r')^2dr'=4 \pi m\left\{\left(\rho_{0} -
  \Frac{\Lambda}{4 \pi G m}\right)\Frac{1}{k^2}\left(\Frac{\sin
    (kr)}{k} - r \cos (kr)\right) + \Frac{\Lambda}{12 \pi G
    m}r^3\right\}.
\ee

 For the distance value $\bar{R}$ for which $\rho_{\!_{\Lambda}}(\bar{R})=0$, we have 
 \be\label{sinkr}
 \Frac{\sin (k\bar{R})}{k\bar{R}}=-\Frac{\Lambda}{(4\pi m \rho_{0} G-\Lambda)}
 \quad .
 \ee
 \noindent For physically relevant results, the right hand side
 (r.h.s.) of (\ref{sinkr}) is supposed to be strictly negative. Hence,
 we have the lower bound for the dark matter density
 $\varrho_{0}=m\rho_{0}> \Frac{\Lambda}{4\pi G}$. An upper bound, of
 course, is obtained when the dark matter density is large enough to
 render the r.h.s. of (\ref{sinkr}) negligible.

 The solution of (\ref{sinkr}) provides the cluster radius with a
 cosmological constant
\begin{equation}\label{rlambda}
 \bar{R}_{\!_{\Lambda}} = 3.19587\times \sqrt{\frac{\hbar^2 a}{Gm^3}}\;. 
\end{equation}

  \noindent The same solution applies for the spin-1 particle, with the
  appropriate substitution of the scattering lengths associated with
  each spin phase.

\section{Statistical analysis}\label{stats}

 In order to constrain the values for the scattering lengths in
 cluster condensates, we use the data obtained from \cite{maughan}
 to construct the Likelihood function $\cal{L}$ for the halo radius $R(a)$ through
\begin{equation}
\label{Like}
{\cal{L}} \propto
\prod_{i=1}^{N} \exp
\left\{  - \frac{1}{2 \, \sigma^2_i}
\left[
R(a) - r_{i}
\right]^2
\right\} \; ,
\end{equation}
 \noindent where $R(a)$ represents the theoretical radius obtained
 from eqs. (\ref{radius0}), (\ref{rpolar}) or (\ref{rlambda}), $r_{i}$
 are the data taken from observations and $\sigma_{i}$ are the errors
 associated with these measurements (not available from
 \cite{maughan}, and therefore overestimated to half the value of each
 measurement). As usual, the maximum value of the probability density
 function derived from (\ref{Like}) gives the best fit value for the
 scattering length parameter $a$. The data set consists of 114 galaxy
 clusters $R_{500}$ radii (the ones which encompass 500 times the
 critical density at each cluster's redshift) in the range
 $0.48-1.91\, {\rm Mpc}$ obtained by X-ray measurements.

\begin{figure}[!htp]
\begin{center}

\includegraphics[scale=0.63]{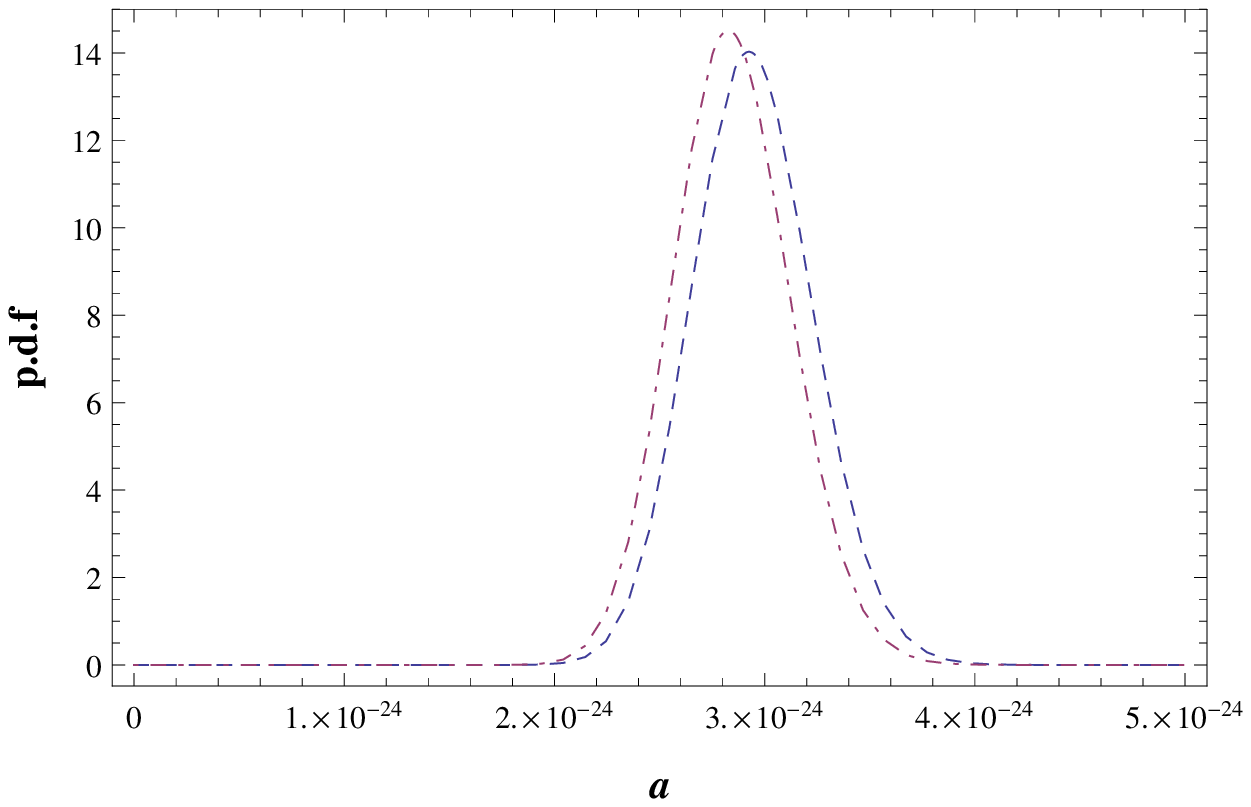}
\hspace{0.1cm}
\includegraphics[scale=0.63]{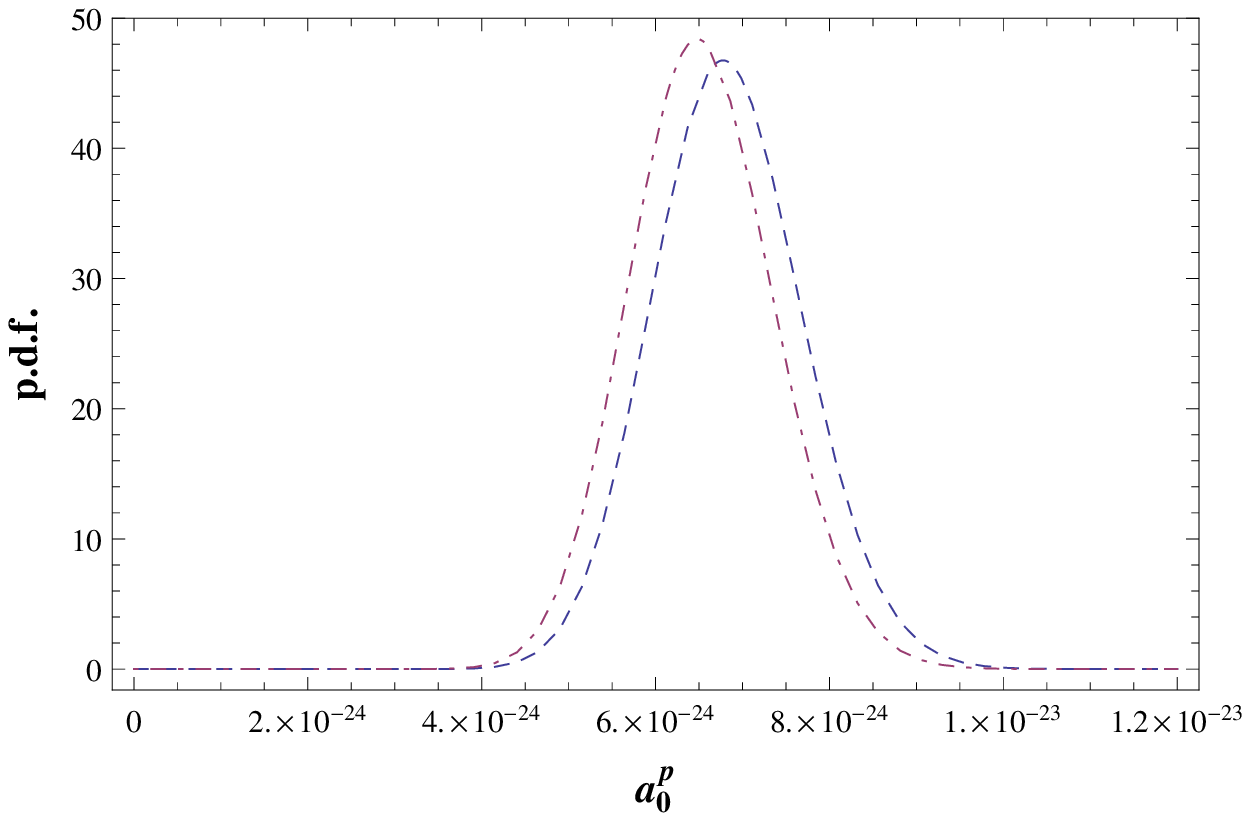}
\caption{Probability density function for the scattering length $a$ of
  a particle with mass $m=10^{-6}\, {\rm eV}$. The left panel refers to a
  spin-0 particle. The right panel refers to a spin-1 particle in the
  polar state. The dashed (dot-dashed) curve in each panel represents
  the non-expanding (expanding) case.}
\label{mass6}

\end{center}
\end{figure}

\begin{figure}[!htp]
\begin{center}

\includegraphics[scale=0.63]{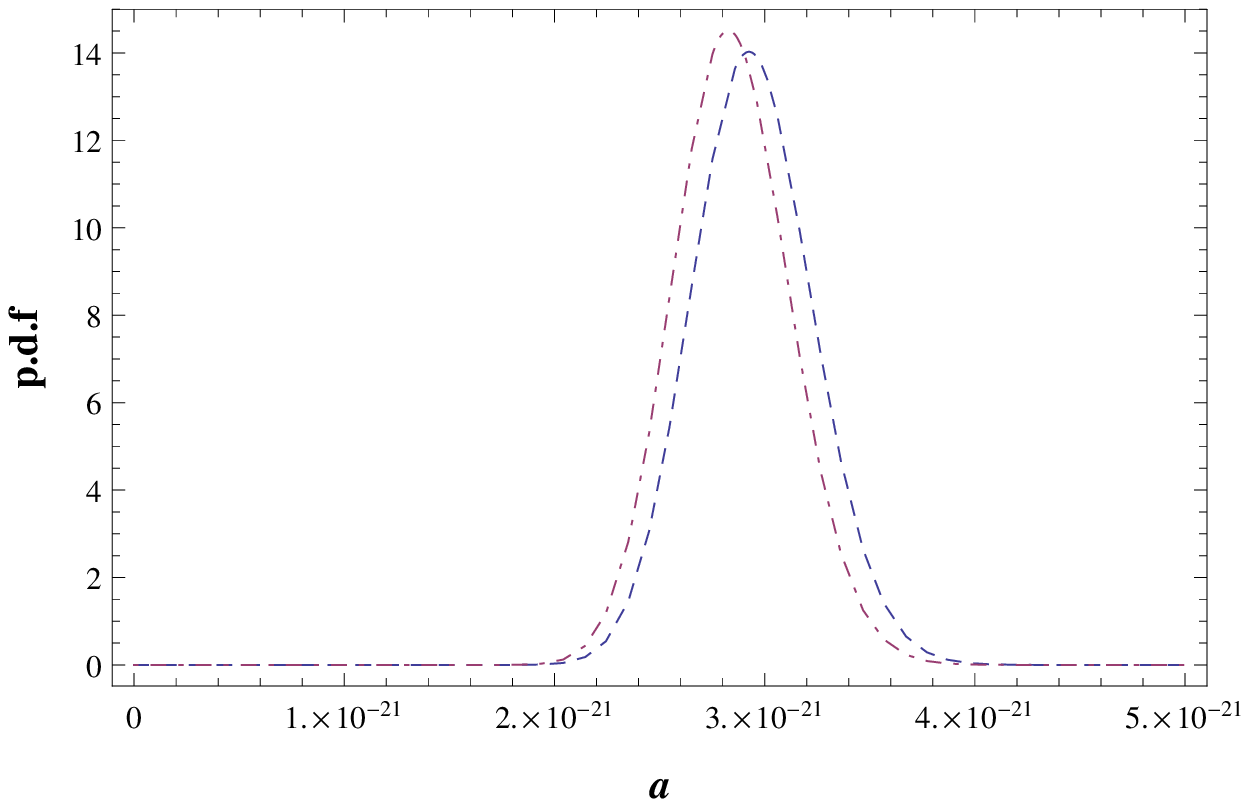}
\hspace{0.1cm}
\includegraphics[scale=0.63]{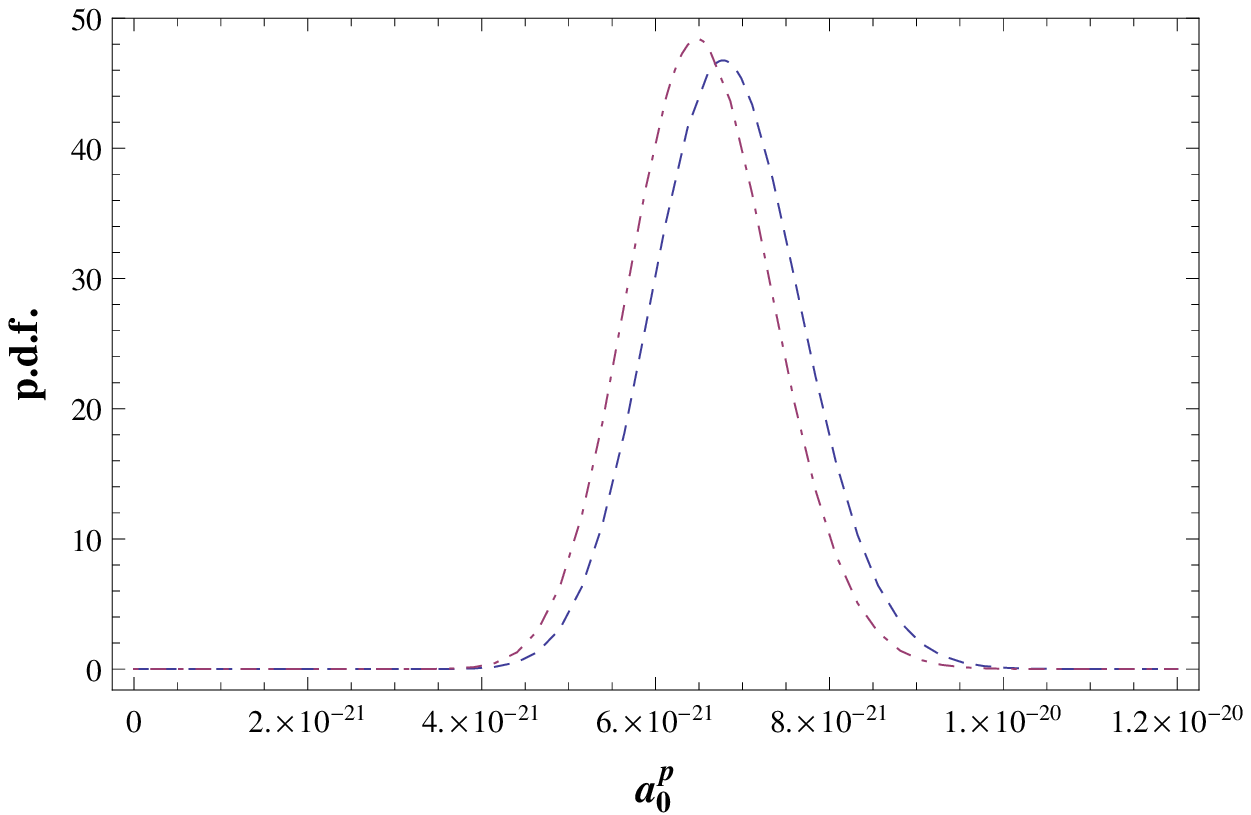}
\caption{Probability density function for the scattering length $a$ of
  a particle with mass $m=10^{-5}\, {\rm eV}$. The left panel refers to a
  spin-0 particle. The right panel refers to a spin-1 particle
  in the polar state. The dashed (dot-dashed) curve in each panel represents
  the non-expanding (expanding) case.}
\label{mass5}

\end{center}
\end{figure}

\begin{figure}[!htp]
\begin{center}

\includegraphics[scale=0.63]{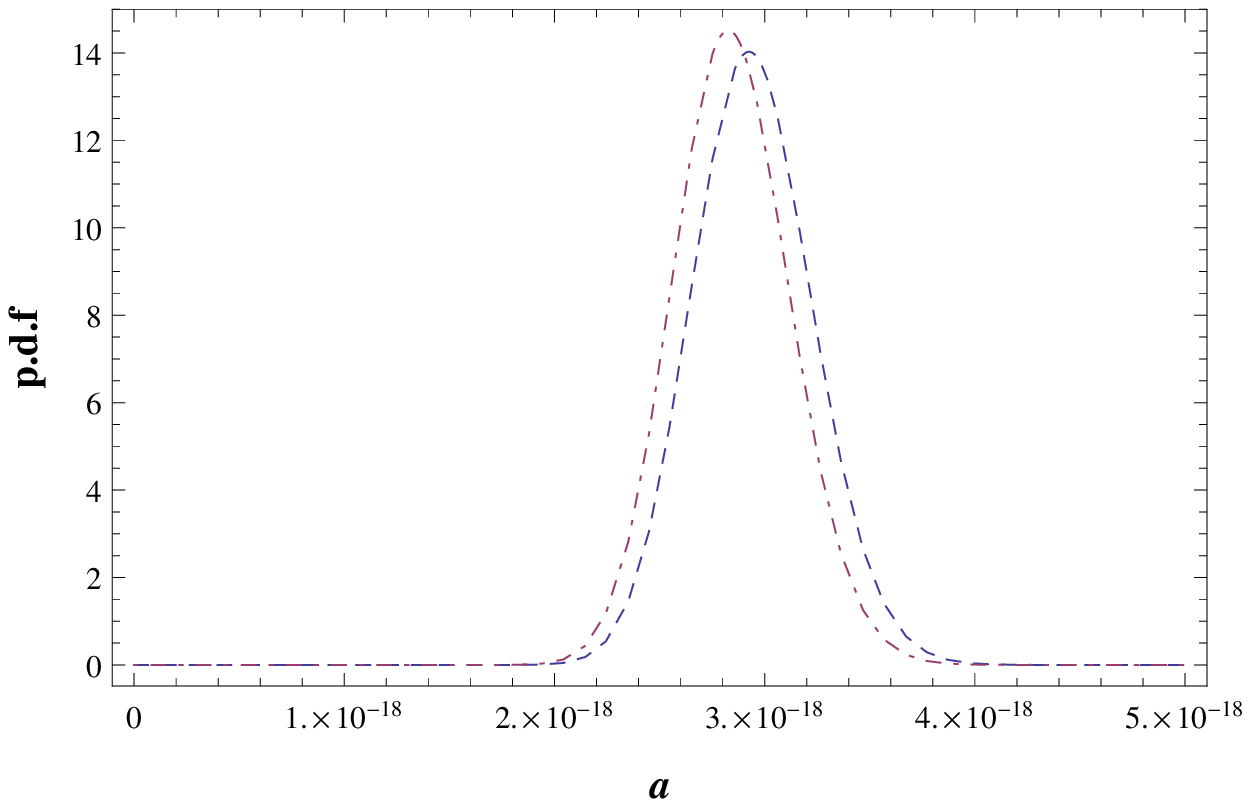}
\hspace{0.1cm}
\includegraphics[scale=0.63]{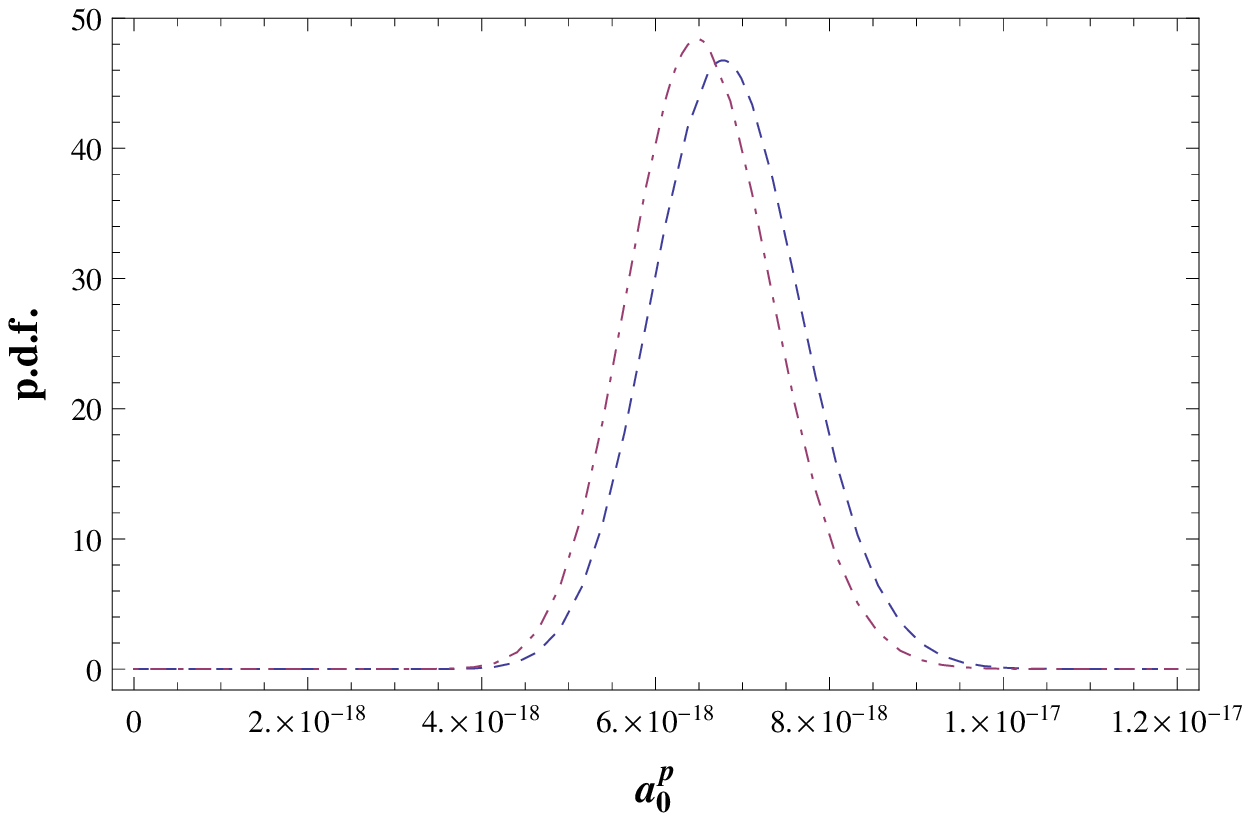}
\caption{Probability density function for the scattering length $a$ of
  a particle with mass $m=10^{-4}\, {\rm eV}$. The left panel refers to a
  spin-0 particle. The right panel refers to a spin-1 particle
  in the polar state. The dashed (dot-dashed) curve in each panel represents
  the non-expanding (expanding) case.}
\label{mass4}

\end{center}
\end{figure}

 Figures \ref{mass6}-\ref{mass4} show the probability density
 functions obtained by this method for the scattering length $a$,
 considering a dark matter particle with mass ranging from $10^{-6}\,
 {\rm eV}$ to $10^{-4}\, {\rm eV}$, and with spin-0 and spin-1. Also, the
 cosmological expansion has been taken into account in the
 calculations. These functions enable us to identify the most probable
 values for the scattering length, given the data set used. We point
 that, for the spin-1 case, we assume that $a_{2}^{f}=a_{2}^{p}=a$,
 and therefore we only constrain the value for $a_{0}^{p}$
 \cite{pir12}.

  Using the previous analysis and the one presented in \cite{pir12},
  we summarize the results obtained for the galactic and the cluster
  condensate in table \ref{tabela-1}. We note that there is a
  difference of five orders of magnitude between the scattering
  lengths, even when the cosmological expansion is taken into account.

\begin{table}\centering
\caption{Values for the scattering lengths for galaxies ($a_{gal}$)
  and clusters ($a_{clu}$) for the particle masses constrained in
  \cite{pir12}.}\label{tabela-1}
\begin{tabular}{ |c|c|c| }
  \hline
  \textit {m} (eV) & $a_{gal}$(m)& $a_{clu}$(m)\\
  \hline
$10^{-6}$ & $10^{-29}$ & $10^{-24}$  \\
  $10^{-5}$ & $10^{-26}$ & $10^{-21}$  \\
  $10^{-4}$ & $10^{-23}$ & $10^{-18}$\\
\hline
\end{tabular}
\end{table}

 Considering the upper bound estimated in \cite{harko12}, which
 implies $a < 10^{-21}\, {\rm m}$, we conclude that the mass value
 $m=10^{-6}\, {\rm eV}$ is favoured in the present analysis. However, we can
 speculate that the significant difference in magnitude between the
 galactic and the cluster cases seems to indicate that the scattering
 length may present some scale dependency, perhaps related to the
 gravitational potential or the total mass of the structure being
 analysed.

 The analysis performed in this section complements the one performed
 in \cite{pir12} for galactic radii.

\section{Orbital velocity}\label{vel_clus}

 In a Newtonian approximation, the attractive force exerted by a large
 scale concentration of mass $M(r)$ on a test particle with mass $m$
 is given simply by 
\be\label{force} F=\Frac{GmM(r)}{r^{2}}, 
\ee
 \noindent which causes the centripetal acceleration on the orbiting
 body.  Using (\ref{force}), the velocity $v(r)=\sqrt{rF/m}$ of the
 test particle around the more massive object (which we can consider
 to be a cluster) is given by
\be\label{velocity}
 v(r)=\sqrt{\Frac{GM(r)}{r}}\, .  
\ee

  In the case of a cluster that is expanding due to a cosmological
  constant $\Lambda$, the centripetal force on the orbiting body is
  written as \cite{nandra}
\be\label{force_lambda}
F=\Frac{GmM(r)}{r^{2}}-\Frac{1}{3}\Lambda m r\, .
\ee
 \noindent From (\ref{force_lambda}), the velocity of a test particle around
 a cluster expanding through the influence of a cosmological constant
 $\Lambda$ is 
\be\label{velocity2}
 v(r)=\sqrt{\Frac{GM(r)}{r}-\Frac{1}{3}\Lambda r^{2}}\, .  
\ee

Substituting the mass function $M(r)$ in (\ref{velocity}) by the one
 obtained from the BH density profile (\ref{psibh}), the orbital
 velocity becomes
\be\label{vel_harko}
 v_{\!_{BH}}(r)=\left[\Frac{4 \pi G \varrho_{\!_{0}}}{k^2}\left(\Frac{\sin (kr)}{kr} -
   \cos (kr)\right)\right]^{1/2}\, . 
\ee

 We can consider the BH density profile modified by the addition of a
 cosmological constant, as shown in equation (\ref{rholambda}). With
 the input of the mass function derived from this profile  in
 (\ref{velocity2}), one obtains
\be\label{vel_lambda} 
v_{\!_{\Lambda}}(r)=\left[4 \pi G m\left\{\left(\rho_{0} -
  \Frac{\Lambda}{4 \pi G m}\right)\Frac{1}{k^2}\left(\Frac{\sin
    (kr)}{kr} - \cos (kr)\right) + \Frac{\Lambda}{12 \pi G
    m}r^2\right\} - \Frac{1}{3}\Lambda r^{2}\right]^{1/2}\, . 
\ee

\begin{figure}[!htp]
\begin{center}

\includegraphics[scale=0.50]{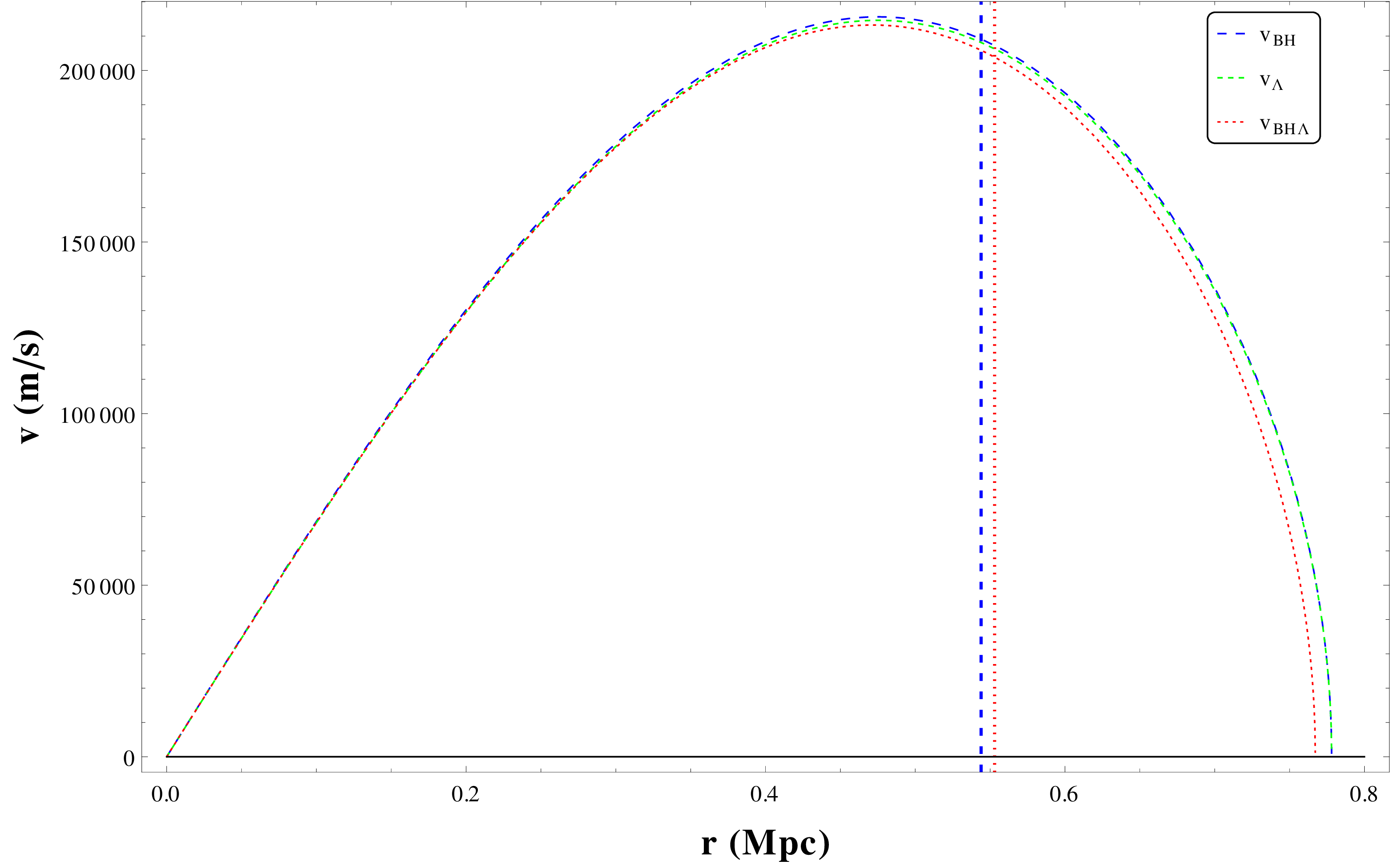}
\caption{Orbital velocity $v$ (in ${\rm m \, s^{-1}}$), of a test particle
  around a cluster, for $m=10^{-6}\, {\rm eV}$. The dashed curve represents
  the non-expanding case, and the dotted curve, the expanding one. The
  vertical lines mark the cluster radii obtained by the expressions
  (\ref{radius0}) and (\ref{rlambda}) (dashed and dotted lines,
  respectively marking $R=0.544\, {\rm Mpc}$ and $R_{\!_{\Lambda}}=0.553\,
  {\rm Mpc}$).}
\label{vel6}

\end{center}
\end{figure}

 As in \cite{nandra}, we can also keep the BH profile unmodified and
 add an expansion term such that the velocity takes the form 
\be\label{vel_bhlambda}
v_{\!_{BH\Lambda}}(r)=\left[\Frac{4 \pi G
    \varrho_{\!_{0}}}{k^2}\left(\Frac{\sin (kr)}{kr} - \cos
  (kr)\right)-\Frac{\Lambda}{3}r^2\right]^{1/2}\, .  
\ee For the case
of a typical cluster, we can use the values $\varrho_{\!_{0}}=10^{-24}\,
{\rm kg/m^3}$, $m=10^{-6}\, {\rm eV}=1.78 \times 10^{-42}\, {\rm kg}$,
$k=1.8\times10^{-22}\, {\rm m^{-1}}$, $\Lambda=1.4\times 10^{-35}\,
{\rm s^{-2}}$ \cite{planck}
to plot the orbital velocity for the test particle, for both the
expanding and non-expanding situations. The plot showing velocity
curves is presented in figure \ref{vel6}.

 In that plot, the point in which the velocity reaches zero value
 corresponds to null centripetal force, i.e., the particle ceases to
 be gravitationally influenced by the cluster and is carried away by
 the cosmological expansion. The vertical lines mark the radii
 obtained by the BH density profile, for the specific parameters
 chosen to draw the figure.

 We can see that $v_{\!_{BH}}(r)$ and  $v_{\!_{\Lambda}}(r)$ present
 no significant differences, specially at large $r$. This is
 consequence of the cancellation of the last two terms in
 (\ref{vel_lambda}). On the other hand, $v_{\!_{BH\Lambda}}(r)$ is notably
 different for larger values of the radius $r$. This is in accordance
 with the findings of \cite{nandra}. Nevertheless, we point out that
 the approximation used in (\ref{vel_bhlambda}) does not include an
 expansion in the mass distribution itself. 

 As expected, the maximum distance $r$ for which the particle is still
 under the force of the cluster is bigger in the case of an expanding
 cluster \cite{nandra}.

\section{Conclusions}\label{conc}
  
 In this work we have considered a Bose-Einstein condensate framework
 for dark matter in clusters. We have assumed an expanding cluster
 embedded in a de Sitter background to include a cosmological constant
 in a simpler Newtonian approximation. This approach allowed us to
 obtain a modified matter density profile for the cluster, slightly
 distinct from the Boehmer-Harko profile in the non-expanding
 situation, and to set the lower bound $\varrho_{\!_{0}}>(\Lambda/4\pi G)$
 for the central dark matter density $\varrho_{\!_{0}}$ with respect to the
 cosmological constant $\Lambda$.

 The cluster radius resulting from this density profile has also been
 derived for both a spin-0 and a spin-1 particle. Using that
 information we were able to use a set of galaxy clusters radii data
 to constrain the scattering length of the Bose-Einstein condensate of
 a particle with masses in the range $10^{-6}-10^{-4}\, {\rm eV}$, in
 a non-expanding as well as in an expanding case (by the inclusion of
 a small cosmological constant in a Newtonian approximation). The
 values obtained for this parameter are typically five orders of
 magnitude larger than what is obtained in the galactic case. This
 result poses the question whether it makes sense to use the same
 value of scattering length parameter for the galactic and the cluster
 scales. Apparently, at least for the mass range we are considering,
 some still undefined scale dependency may take place to describe the
 condensate in galaxies and clusters by using its microscopic
 parameters. Another speculative possibility is the dominance of a
 different kind of dark matter fluid in clusters, composed of
 particles endowed with distinct parameters from the ones which form
 galactic haloes. This possibility could be further explored in the
 mixed dark matter scenario \cite{medvedev14}.

 We have used the Newtonian approximation for the gravitationally
 bound system in order to calculate the orbital velocity of a test
 particle around an expanding cluster. The maximal radius in which the
 velocity is null sets the greatest distance of influence of the dark
 matter cluster. This distance shows no appreciable difference in
 comparison to the non-expanding case when the new derived density
 profile is used, due to the smallness of the cosmological
 constant. This result is in contrast with the one found in
 \cite{nandra}, which did not consider a modified density profile for
 the cluster.

 The issue of the most adequate values of the parameters of condensate
 dark matter in large scale structures has been occupying a
 considerable space in the literature, and we hope that the
 considerations presented here may guide us in future works on this
 subject.
 
\begin{acknowledgments}

 The authors acknowledge CAPES (Coordena\c c\~ao de Aperfei\c coamento
 de Pessoal de N\'\i vel Superior) for financial support and M. Pires
 for useful discussions. The authors are also grateful to an
 anonymous referee for helpful suggestions.

\end{acknowledgments}

\end{document}